# DESCRIBING THE FLOW CURVE OF SHEAR-BANDING FLUIDS THROUGH A STRUCTURAL MINIMAL MODEL

Daniel Quemada [1,*] and Claudio Berli [2]

[1] *Groupe de Rhéologie, Matière et Systèmes Complexes, Univ. Paris-Diderot (Paris 7), Bât. Condorcet, MSC, case 7056 - 75205 Paris Cedex 13, France.*
[2] *INTEC (UNL-CONICET), Güemes 3450, 3000, Santa Fe, Argentina, and Dpto. de Física, FBCB, UNL, Ciudad Universitaria, 3000, Santa Fe, Argentina.*

*E-mail: danielquemada@orange.fr

**Abstract:**
Main characteristics of colloidal systems that develop fluid phases with different mechanical properties, namely shear-banding fluids, are briefly reviewed both from experimental and theoretical (modelling) point of view. A non-monotonic shear stress *vs.* shear rate constitutive relation is presented. This relation derives from a phenomenological model of a shear rate-dependent viscosity describing structural changes and involves the possibility of multivalued shear rates under a given shear stress. In the case of a stress-dependent viscosity, the same model allows one to predict vorticity banding. Predictions of this model under controlled stress are discussed, namely occurrence of a kind of top- and bottom-jumping of the shear rate in response to stress increasing-decreasing. Applying this model to evaluation of the flow curve of such colloidal systems is performed. Particular emphasis is placed on the adequate computation of the shear rate function in cylindrical Couette cells in order to handle the corresponding flow curve which exhibits the well-known shear stress plateau. Indeed, as different fluid phases coexist in the flow domain, measured (torque *vs.* angular velocity) data cannot be directly converted into rheometric (shear stress *vs.* shear rate) functions. As the lacking non-local terms in the model prevents the direct determination of the stress-plateau, this value is included as an adjustable parameter. Thus model predictions satisfactorily match up experimental data of wormlike micellar solutions from the literature.

**Keywords:** Shear-banding, Couette rheometry, wormlike micellar solutions, structural model.

# 1 - Introduction

Under shear flow conditions, many complex fluids develop fluid phases with different mechanical properties, namely shear bands, parallel to the flow direction, where the shear rate takes different mean values. This so-called shear banding behaviour has been interpreted as resulting from flow instabilities associated to the existence of a non-monotonic flow curve with a region of negative slope. Under controlled stress experiments, band coexistence occurs in a certain range of shear rate where the flow curve presents either a stress plateau or a notably reduced slope. Typical results found in the analysis of wormlike micellar solutions (WMS) are represented in Fig.1.

Shear banding has mainly been found experimentally in WMS [1−6], colloidal crystals [8−10] and lamellar surfactant systems [11,12]. In these materials, shear-induced changes of the microstructure have been observed by birefringence [4,6,12], local light and neutron scattering [13,14], associated with various methods of velocimetry [7,12,15−18], performed in various geometries (pipe, cone-and-plate and coaxial cylinders). In many cases, experimental findings prove that banding can be interpreted as a I-N phase transition between a disordered (Isotropic) phase coexisting with an ordered (Nematic) one. Hence the fluid exhibits shear-thinning, with low viscosity as micelles align in the flow direction at high shear rate. By contrast, micelle entanglements lead to high viscosity at low shear rate. In Fig.1, inset boxes schematically show the microstructural changes driven by shear flow. General agreement between birefringence and velocity profiles have been observed, although some systems exhibit a high viscous nematic state at high shear rate, which could reveal either a possible mesoscale ordering [16] or shear-thickening effects as yet observed in similar systems in this rate domain [19].

Another kind of banding formation is called *vorticity banding* because band interfaces appear as turbid and clear (pancake-like) rings oriented perpendicular to the vorticity axis. However no steady state has been observed in this case due to stress oscillations that take place once the pancake structure appears [20,21,22].

A large number of theoretical works have demonstrated that the main feature in shear banding modelling is the need of a non-monotonic constitutive relation in order to obtain a flow curve, shear stress $\sigma$ *vs*. shear rate $\dot{\gamma}$, which exhibits a region where the slope $d\sigma/d\dot{\gamma}$ is negative.

Basically, it is admitted that the flow curve should take the form shown in Fig.2, with maximum and minimum shear stresses, $\sigma_M$ and $\sigma_m$, reached at shear rates $\dot{\gamma}_M$ and $\dot{\gamma}_m$, respectively. For any given stress $\sigma_0$ lying within the interval [$\sigma_m$, $\sigma_M$], one may define two "stable" shear rates, $\dot{\gamma}_1$ and $\dot{\gamma}_2$, the intermediate solution $\dot{\gamma}_3$ corresponding to $d\sigma/d\dot{\gamma} < 0$, hence to an instable state. For simple shear in planar Couette, force balance ($\nabla \cdot \sigma = 0$) requires that $\sigma$ remains uniform across the gap, $\sigma_0 = \sigma(\dot{\gamma}_1) = \sigma(\dot{\gamma}_2)$, which suggests that the flow is shared out in two coexisting bands with *constant* rates $\dot{\gamma}_1$ and $\dot{\gamma}_2$. The average shear rate thus follows a lever rule $\dot{\gamma}_{av} = x\dot{\gamma}_1 + (1-x)\dot{\gamma}_2$, where $x$ gives the interface location within the gap of the rheometric cell, with $0 < x < 1$ [23]. All streamlines being equivalent in steady planar Couette flow, the presence of shear bands should be possible for any stress $\sigma_0$ such as $\sigma_m < \sigma_0 < \sigma_M$, thus band interface location cannot be determined. However, experiments in real systems have shown the robustness of a well defined stress plateau, $\sigma = \sigma^*$, at which the two phases coexist. Some criterion is required for selecting this value. Therefore, modelling should incorporate some new ingredient. Note that in the case of non-planar geometries, the stress variation within the gap cannot give the selection criterion, although it might explain the experimental observation of a quasi-plateau in this curve instead of a true one.

Although the way of spontaneous formation of a banded flow still remains an open question, several works have proposed that instabilities should result from a flow/structure coupling, *i.e.* coupling between shear induced changes of the internal structure of the material (as the isotropic-to-nematic transition in WMS) and pure hydrodynamic instabilities associated to the presence of bands which induce non-homogeneous velocity profile [1,20].

The basic role played by molecular orientation in I-N transition and more generally in nematic structures has been treated by Doi and Hess theories which appear as the more powerfull tools to solve this problem [24–26], leading to applications among which number of works should be mentionned [27–31].

In addition, many situations exist where the observed unsteady properties result in propagating waves or instabilities, leading to periodic motions (stick-slip like, alternating bands…) or irregular chaotic behaviour [22,29–31,36–39].

On the other hand, it was early suggested from instabilities and hysteresis cycles observed in polymer melt extrusion that assuming a multivalued flow curve could be sufficient to generate instability [40,41]. More recently, extensive work on stability analysis has confirmed that the region where $d\sigma/d\dot{\gamma} < 0$ is indeed responsible of band formation [32–35,42]. However, most of these theoretical models fail in predicting the robustness of the stress value at the plateau, $\sigma^*$, if they do not incorporate in the constitutive relation some non-local effect as diffusion terms concerning either concentration [43,44], shear stress [42,43,45–47], shear rate [35] or both shear stress and shear rate [48]. This allows to change each (unphysical) discontinuity in $\dot\gamma$ into a "material interface", the thickness of which is of the order of the diffusion length, with $\dot\gamma$ varying continuously within the interface from low to high values in outside bands, $\dot\gamma_1$ and $\dot\gamma_2$. Note that considering only a coupling through concentration-dependent stress $\sigma(\phi)$ − or viscosity $\eta(\phi)$ − where $\phi$ is the particle volume fraction also allowed to obtain a selection of $\sigma^*$ as in reaction-diffusion models [49].

As a contribution to phenomenological modelling of shear banding, the present work is a first step of a simple approach without including in its present form such non-local effects, hence discarding any possibility to predict both band formation and plateau location. Based on a simple model for the local shear viscosity $\eta(\dot\gamma, \phi)$ which allows a multivalued flow curve, this work is only devoted to discuss (i) some characteristics of this model under shear banding conditions, (ii) rheological aspects concerning the general treatment of experimental data of shear banded flows in cylindrical Couette cells and (iii) the application of the resulting method to some actual rheometrical data.

This paper is organized as follows. In Sec.2, the Structural Minimal model (hereafter called SM-model for convenience) based on a *structural viscosity* is presented, underlining main specific features in comparison with several minimal models recently proposed in the literature. SM-model characteristics for shear-banding under steady conditions are discussed in Sec.3. After a brief review of the problem related to the determination of the shear rate function in monophasic flows, special attention is devoted to the calculation of the torque *vs.* angular velocity under stress-banding conditions, in order to predict more precisely actual rheometrical data in a cylindrical Couette cell. This is successfully achieved in Sec.4 using the SM-model, allowing to discuss in Sec.5 an example of application to experimental data of WMS. Finally, some concluding remarks and comments on future development of this work are given in Sec.6.

## 2 – The SM-model

### 2.1 – Foundations of the viscosity equation

The SM-model has been deduced from minimization of viscous energy dissipation in concentrated suspensions [50] that led to a (Krieger-like) viscosity-volume fraction relation, $\eta(\phi) \sim (1-\phi/\phi_m)^{-2}$. This relation traduces a transition-like behaviour close to maximum packing of particles, $\phi_m$, with a "critical exponent" *2*. For dispersions of aggregating particles, forming "Structural Units" (SU), the SU-concept led to generalize this relation to the form $\eta(\phi_{eff}) \sim (1-\phi_{eff}/\phi_m)^{-2}$, where $\phi_{eff} =(1+CS)\phi$ is the effective volume fraction of the disperse phase, *S* being the fraction of individual particles within all of SU, and *C* a factor depending on the mean compactness of SU (see [51] for further details on these definitions).

As a structural variable, *S* takes into account the volume of suspending fluid immobilized within SU [51,52]. Finally, shear-induced structural changes within the system are incorporated into the model assuming that *S* obeys a kinetic equation [53]. It is assumed that, when a given shear rate is applied, a sort of order-disorder equilibrium is established, in which the forward (Brownian motion) and backward (shear-induced ordering) processes balance. This leads to the following form for the *structural viscosity*

$$\eta = \eta_\infty \left[ 1-(1-\chi)S \right]^{-2} \qquad (1)$$

In Eq.(1) $\eta$ is a function of $S=S(\phi, \Gamma, t)$, where $\Gamma = \dot{\gamma}/\dot{\gamma}_C$ is a reduced shear rate, $\dot{\gamma}_C$ is a characteristic shear rate ($\approx$ reciprocal of a structural relaxation time $t_C$) and $\chi = (\eta_\infty/\eta_0)^{1/2}$, $\eta_\infty$ and $\eta_0$ being the limiting viscosities at $\Gamma >> 1$ and $\Gamma << 1$, respectively. Note that these limits correspond to complete rupture and complete building of the structure, taking the corresponding values $S_\infty=0$ and $S_0=1$ for simplicity (see [53]).

Applying such a phenomenological model to systems like WMS is tentatively made in this work. As this model has been successfully used to describe a very large variety of concentrated dispersions [54], it is thought that some details of structural processes at the microscale would have a weak influence on rheological properties at the mesoscale. For instance, in the special case of WMS, details on reversible scission and recombination of chains and/or reptation process in the entangle state will be forgotten, keeping only the following picture : (i) the limiting structure at low shear rate is formed by large entangled groups of disordered micelles which immobilize a large amount of suspending fluid, thus having a high viscosity $\eta_0$ ; (ii) as the shear rate increases, these groups are progressively broken down with releasing some part of the entrapped fluid; (iii) finally, at high shear, both micelle alignment in the flow direction and complete release of entrapped fluid lead to a small value for $\eta_\infty$. It is expected that, despite this approach remains approximate, it benefits from a very limited number of parameters, moreover having a clear physical significance.

## 2.2 – Prediction of shear-banding

Under steady conditions, $\Gamma = Cte$, the shear stress variations are given by

$$\sigma(\dot{\gamma}) = \eta\dot{\gamma} = \eta_\infty \dot{\gamma}_C \left(\frac{1+\Gamma}{\chi+\Gamma}\right)^2 \Gamma \tag{2}$$

This model exhibits a non-monotonic flow curve $\sigma(\Gamma)$ within a range of $\chi$-values, hence leads to a multivalued shear rate for a given stress. Here we will use the variable $\alpha = \eta_0/\eta_\infty = \chi^{-2}$ for convenience. Fig.3A displays these variations in terms of a reduced stress $\Sigma = \sigma/\sigma_C$ vs. $\Gamma$ (taking $\sigma_C = \eta_\infty \dot{\gamma}_C$ for the sake of simplicity, without reducing the generality of the results). As yet underlined above, Eq.(2) defines a (local) reduced steady viscosity $H = \eta/\eta_\infty = H(\Gamma, \alpha)$ as an explicit function (i) of shear rate $\dot{\gamma}$ through the reduced variable $\Gamma$, (ii) of volume fraction through the $\phi$–dependent ratio of limiting viscosities, $\alpha = \eta_0/\eta_\infty$. Non-monotonicity of flow curves $\Sigma(\Gamma)$ occurs for $\alpha > 81$, the higher the $\alpha$-value, the more pronounced the non-monotonicity.

It is interesting to observe however that the viscosity $\eta(\dot{\gamma}) = \sigma(\dot{\gamma})/\dot{\gamma}$ related to Eq.(2) is a purely monotonic function, even within the region of multivalued shear rate for a given stress. In terms of the reduced viscosity $H(\Gamma)$, Fig.3B shows behaviours that are always monotonic: *shear thinning* if $\alpha > 1$ (thus specially in the multivalued region) and a *shear thickening* if $\alpha < 1$.

## 2.3 – Prediction of vorticity-banding

An interesting feature of this model is the possibility to use it for modelling vorticity banding. Indeed, if the steady viscosity explicitly depends on the stress, the flow curve is defined by $\dot{\gamma} = \sigma/\eta(\sigma)$. In this case, one should use a reduced shear stress $\Sigma = \sigma/\sigma_C$ instead of the reduced shear rate $\Gamma$ in Eq.(2), with a characteristic shear stress $\sigma_C$ and keeping $\Gamma = \dot{\gamma}/\dot{\gamma}_C$ where $\dot{\gamma}_C = \sigma_C/\eta_\infty$ (*i.e.* still with $\sigma_C = \eta_\infty \dot{\gamma}_C$), that leads to the flow curve in reduced variables,

$$\Sigma = [(1+\Sigma)/(\chi+\Sigma)]^2 \Gamma \tag{3}$$

Figure 4A displays the reduced flow curves $\Sigma = \Sigma(\Gamma)$ for different values of $\beta = \eta_\infty/\eta_0 = \chi^2$. A domain of multivalued stress at given shear rate is observed for $\beta > 81$. Corresponding curves for the reduced steady viscosity as a function of $\Sigma$

$$H(\Sigma) = \eta/\eta_\infty = [(1+\Sigma)/(\chi+\Sigma)]^2 \tag{4}$$

are shown on Fig.4B. By contrast with the shear-banding case, this multivalued domain corresponds to a shear thickening behaviour. Note that in terms of shear rate the reduced viscosity $H(\Gamma)$ exhibits on Fig.4C a multivalued domain for $\beta > 81$ as expected since a multivalued stress is then associated to a given $\Gamma$. Similar findings would be obtained for $H(\Sigma)$ in the shear banding case.

**2.4 – Comparison with several minimal models**

SM-model can be compared to other phenomenological ones recently proposed in the literature. In particular, the requirement for a multivalued flow curve has led to *ad-hoc* models, often called "minimal models", the quality of them being their simple analytical form that allows to perform stability analysis of band formation. For instance, still keeping the notation $\Gamma$ for the reduced shear rate, the simplest (trinomial) dependence under steady conditions $\sigma \sim \dot{\gamma} \eta_0/(1+\Gamma^2)$ was taken for either the total shear stress (Eq.(9) in [55]) or its viscoelastic part only (Eqs.(6) and (7) in [42]). A little less simple expression (also depending on $\eta_\infty$),

$$\sigma \sim \dot{\gamma}\frac{\eta_0 + \eta_\infty \Gamma^2}{1+\Gamma^2}, \tag{5}$$

has been used (Eq.(8) in [55] and Eq.(33) in [35]). The nonlinearity in Eq.(5), resulting from the shear rate dependence in $\Gamma^2$, leads automatically to a trinomial equation under controlled stress condition, hence to the possibility of having 3 roots. It is noteworthy that there is not other justification for selecting this kind of expression among the simplest ones. By contrast, the nonlinearity in Eq.(2) results from the nonlinear $\phi$-dependence of $\eta$, $\eta \sim (1-\phi/\phi_m)^{-2}$, even (as it is considered in this paper) if the shear rate dependence of the ratio of kinetic constants remains linear (see [53]). Therefore, using Eq.(2) appears less arbitrary than using Eq.(5).

Very similar comments can be given in the case of vorticity banding, with a constitutive equation containing an *ad-hoc* third order polynomial $R(\sigma)$, leading to a trinomial equation in $\sigma$ under controlled shear rate condition (see for instance Eqs.(1) and (2) in [47], discarding time and non-local effects, *i.e.* in steady homogeneous conditions). Such an equation corresponds to the one deduced from the relation $\sigma/\eta(\sigma) = \dot{\gamma}$ with $\eta(\sigma)/\eta_\infty$ given by Eq.(4).

## 3 - SM-Model characteristics under steady shear-banding conditions.

In order to predict the stress plateau associated to shear banding, the *top-jumping* concept have been at first introduced. Although this was done on the basis of satisfactory agreement between model predictions [41,56] and WMS data [57], stress startup behaviour under constant $\dot{\gamma}$ shows clearly that flow curve determination requires to perform measurements exceedingly slowly. Transient stress responses obtained using startup experiments showed indeed two different behaviours. Below the plateau regime ($\dot{\gamma} < \dot{\gamma}_1$) the stress grows rapidly (in times ≤ Maxwell time $\tau_M$) up to a stationary value and remains constant. Above $\dot{\gamma}_1$, in the plateau region, the transient stress exhibits the same rapid increase at short times followed by a slow relaxation on very longer times $\tau_R >> \tau_M$, fairly described by a stretched exponential [5,33]. Such findings have demonstrated that reaching steady conditions requires very long delay in the plateau regime, likely due to the development of structures associated with band formation. Therefore, it appears that number of "steady" measurements, which under (up and down) imposed stress yield flow curves similar to hysteresis cycles [59,60], actually correspond to unsteady behaviour. Indeed, a flow-curve resembling top-jumping can be recorded by fast stress-controlled scans (roughly one data point per time ≈ $10\tau_M$) [2]. In some extent, the SM-model may predict such a unsteady behaviour, as follows.

Considering a fluid governed by Eq.(2) with $\chi < 9$, one may discuss the predicted shear rate if the stress is increased fast enough (with a scanning time $\tau_S$ such as $\tau_M < \tau_S << \tau_R$) to assume that structural changes are delayed, i.e. keeping approximatively the viscosity value of the previous step. Such an assumption is acceptable if $\tau_S$ is of same order than $\dot{\gamma}_C^{-1}$, the SM-model characteristic time for structural changes, that seems plausible from startup experiments. Hence, for a given set of successive stress values $\sigma_n$, the corresponding set of shear rate values are $\dot{\gamma}_n = \sigma_n / \eta(\dot{\gamma}_{n-1})$, where the subindex $n$ stands for the $n$-th data point.

Fig.5A displays the resulting variations of $\Sigma$ *vs.* $\Gamma$ (same variables yet introduced in the previous section) under controlled stress conditions. A *top-jumping*-like behaviour is clearly observed after reaching the flow curve maximum $\Sigma_{max}$. However, as shown on Fig.5B, details of variations $\Sigma(\Gamma)$ close to the maximum of the flow curve show two main characteristics :
(i) $\Sigma$ reaches progressively the apparent plateau, with a "$\Sigma_{jump}$-value" clearly above $\Sigma_{max}$
(ii) the successive data points in response to identical stress increments $d\Sigma$ correspond to a series $\Gamma_n$ which exhibits successive jumps with an increasing increment $d\Gamma_n = (\Gamma_n - \Gamma_{n-1})$.

These predictions are in fair agreement with several experiments (see Figs.3a-7 in [2]) which exhibit both the smooth transition to an apparent plateau (well above the "true" plateau for shear banding) and data points with increasing $d\Gamma$. This confirms that stress controlled flow curves displayed on Fig.5 are definitely nonequilibrium curves and concern successive structural states which could be considered as points of some "metastable extension" of the low shear monophasic branch.

Lastly, Fig.6 shows that decreasing $\Sigma$ from the high shear branch of the flow curve leads to a *bottom-jumping*-like behaviour hence, with the previous $\Sigma$-increase, to the formation of an hysteresis cycle very similar to the ones observed experimentally [59,60].

As yet discussed in Sec.1, top or bottom jumpings thus do not give the good criterion for selecting $\sigma*$. This can be only achieved by using theories which involve constitutive equations with nonlocal terms. Therefore, we do not proceed to further development of the SM-model before introduction of such non-local effects. In the next sections, the discussion will be limited to the applicability of this model to the prediction of some data.

## 4- Equations of Couette rheometry

After a brief review of the problem related to the determination of the shear rate function in monophasic flows, a special attention is devoted to the calculation of the torque *vs*. angular velocity under stress-banding conditions, in order to predict more precisely actual rheometrical data in a cylindrical Couette cell. This is successfully achieved using the SM-model.

### 4.1 - Monophasic flows

Determining the flow curve $\sigma(\dot{\gamma})$ of non-Newtonian fluids requires the knowledge of both the shear stress $\sigma$ and the shear rate $\dot{\gamma}$ at one place in the flow domain of the cell. In rheometry between concentric cylinders (discarding end effects), it is worth to observe that velocity vector $\underline{v}$ and shear stress tensor $\underline{\underline{\sigma}}$ in cylindrical coordinates [$r, \theta, z$] reduce [61] to their components $v_\theta = v(r,t)$ and $\sigma_{r\theta} = \sigma(r,t)$, leading to the balance condition $\sigma(r,t) = \eta\,\dot{\gamma}(r,t)$, where $\eta$ is the shear viscosity and $\dot{\gamma}$ the shear rate function, defined as $\dot{\gamma}(r,t) = r\,\partial(v/r)/\partial r$. Under steady conditions, the functions of interest, $\sigma$ and $\dot{\gamma}$, are related to measured quantities, angular velocity $\Omega$ and torque $M$, through the following equations (for example, [62,54]),

$$\sigma(r) = \frac{M}{2\pi r^2 L}, \qquad (6)$$

$$\Omega = \int_{\kappa R}^{R} \frac{\dot{\gamma}(r)}{r} dr, \tag{7}$$

where $R$ and $\kappa R$ are the radii of outer and inner cylinder, respectively ($0 < \kappa < 1$), and $L$ is the cylinders height. While $\sigma(r)$ is obtained straightforwardly from Eq. (1), extracting $\dot{\gamma}(r)$ requires inverting the integral of Eq.(7), the solution of which is not unique due to the scattering present in experimental data. This inverse problem is also designated ill-posed in the literature [64,65]. Indeed, direct estimations of $\dot{\gamma}$ can be achieved only when the gap between inner and outer cylinders is very small, say $0.99 < \kappa < 1$. Under these circumstances, $\dot{\gamma}(r)$ is considered nearly uniform throughout the flow domain, and thus Eq.(7) can be replaced by $\dot{\gamma}_{ng} \approx \Omega \kappa/(1-\kappa)$, which is known as the narrow-gap solution for Couette viscometry [62,54].

For wider gap widths, an approach commonly used in practice consists in estimating the shear rate at the inner wall, $\dot{\gamma}_{\kappa R} = \dot{\gamma}(\kappa R)$, by using the following asymptotic expression [66],

$$\dot{\gamma}_{\kappa R} = \frac{\Omega}{ln(1/\kappa)} + \frac{d\Omega}{d\,ln\,\sigma_{\kappa R}} + \frac{ln(1/\kappa)}{3} \frac{d^2\Omega}{d\,ln\,\sigma_{\kappa R}^2} + ..., \tag{8}$$

where $\sigma_{\kappa R} = M/(2\pi\kappa^2 R^2 L)$ is the shear stress at the inner cylinder. This equation is recommended in rheology books (for instance, [54]) to be used for $0.5 < \kappa < 1$. It is relevant to mention that the error in terminating the series at the second term is lower than $[ln(1/\kappa)]^2$. As an interesting alternative to estimate $\dot{\gamma}_{\kappa R}$, finite difference formulas that avoid the numerical differentiation of raw data were recently proposed [67].

### 4.2 - Shear-banding flows

Shear-banding flows involve an additional complexity: abrupt changes in the function $\dot{\gamma}(r)$ arise when two or more phases coexist in the flow domain of the cell. As described in the introduction, different theoretical models aimed to predict the occurrence of shear-banding, with a general consensus that the underlying constitutive curve of shear-banding fluids has the form shown in Fig.2.

Although the observed robust selection of the stress plateau $\sigma = \sigma *$ is only obtained with theories including non-local (diffusive) terms in the constitutive model, we intend to show that it is possible to limit the modelling of stress controlled measurements to these outside bands, considering the plateau value $\sigma^*$ as a model variable to be adjusted together with the SM-model parameters, $\eta_\infty$, $\eta_0$ and $t_C \equiv \dot{\gamma}_C^{-1}$.

In Couette cells, the shear stress is maximum at the inner cylinder ($\sigma_{\kappa R}$), and decreases smoothly as $\sigma(r) = \sigma_{\kappa R}(\kappa R/r)^2$, to give the minimum value at the outer cylinder ($\sigma_R = \kappa^2 \sigma_{\kappa R}$). The simple (generally accepted) scenario considered here is that, when the stress $\sigma_{\kappa R}$ reaches the value $\sigma^*$, a new phase with lower flow resistance develops from the inner cylinder, and consequently $\dot{\gamma}_{\kappa R}$ jumps from $\dot{\gamma}_1$ to $\dot{\gamma}_2$ (Fig. 2). Further increase in $\sigma_{\kappa R}$ results in the growth of this new phase, with a thickness enhancement of the associated band, up to complete development of the same phase in the whole gap, as $\sigma_{\kappa R}$ reaches $\sigma^*/\kappa^2$.

For the purpose of describing our rheometric problem, we start with the following assumptions

(i) shear-bands are stable and can coexist in steady state conditions;
(ii) the bands present a localized, flat and sufficiently narrow interface at a certain $r = r^*$, where the shear stress is $\sigma = \sigma^*$;
(iii) both shear stress $\sigma(r)$ and fluid velocity $u(r)$ are continuous at the interface.

In principle, these assumptions agree with hypothesis and experimental data reported in the literature ([6,7,56,68] for instance). However, in relation with the interface (ii), very recent theoretical [46] and experimental [69] works suggest the existence of a region of instability between the bands in the vorticity direction. As a first approximation, here we assume a sufficiently narrow and flat interface between the bands.

Under these conditions, and considering the SM-model, Eq.(2), defining $\sigma(\dot{\gamma})$ as that plotted in Fig. 2, the expressions of $\Omega$ as a function of $\sigma_{\kappa R}$ in the different stress domains are given as follows (see also [68]),

$$\sigma_{\kappa R} \leq \sigma^*, \qquad \Omega = -\frac{1}{2}\int_{\sigma_{\kappa R}}^{\sigma_R} \frac{\dot{\gamma}^<(\sigma)}{\sigma}d\sigma; \qquad (9a)$$

$$\sigma_{\kappa R} \geq \sigma^* \geq \sigma_R, \qquad \Omega = -\frac{1}{2}\left\{\int_{\sigma_{\kappa R}}^{\sigma^*} \frac{\dot{\gamma}^<(\sigma)}{\sigma}d\sigma + \int_{\sigma^*}^{\sigma_R} \frac{\dot{\gamma}^>(\sigma)}{\sigma}d\sigma\right\}; \qquad (9a)$$

$$\sigma_R \geq \sigma^*, \qquad \Omega = -\frac{1}{2}\int_{\sigma_{\kappa R}}^{\sigma_R} \frac{\dot{\gamma}^>(\sigma)}{\sigma}d\sigma. \qquad (9c)$$

In these equations, $\dot{\gamma}^<$ and $\dot{\gamma}^>$ indicate shear rate values $\dot{\gamma} < \dot{\gamma}_1$ and $\dot{\gamma} > \dot{\gamma}_2$, respectively. To handle the inverse problem given by Eqs.(9a)-(9c), we need to introduce the constitutive equation $\sigma(\dot{\gamma})$ characterizing the fluid under study.

## 4.3 - Calculations proposed for shear-banding flows

### 4.3.1 - Direct and inverse calculations

For a given fluid, if the parameters ($\eta_0$, $\eta_\infty$, $t_c$) entering the SM-model are known, Eqs. (2) and (9a)-(9c) predict the values $\Omega$ *vs.* $\sigma$ (or *M*) to be obtained in an experiment in which the requirements to accomplish a viscometric flow are satisfied (steady state, no-slip at the walls, end effects negligible, isothermal flow), as well as the assumptions (i-iii) made above. This possibility, designated 'direct calculation', will be illustrated in detail in Section 5.

On the other hand, a more challenging problem is the inverse calculation: determining the values of the parameters *($\eta_0$, $\eta_\infty$, $t_c$)* from the curve of raw data $\Omega$ *vs.* $\sigma$ (or *M*), and then using them to plot the flow curve $\sigma(\dot{\gamma})$ in the appropriate range of shear rates. The implementation of this task requires further efforts, because the minimization problem involved cannot be solved with standard mathematical software (for instance, [70]).

### 4.3.2 - Numerical procedure for the direct calculation

The values $\dot{\gamma}^<(\sigma)$ and $\dot{\gamma}^>(\sigma)$ entering Eqs.(9a)-(9c) are obtained from Eq.(2), as numerical roots $\dot{\gamma}(\sigma)$ for a given set of known parameters ($\eta_0$, $\eta_\infty$, $t_c$). This task is carried out through a Newton-Raphson subroutine [71], for around $10^4$ discrete values of $\sigma$ in the range $\sigma_R \leq \sigma \leq \sigma_{\kappa R}$. Then Eqs.(9a)-(9c) are integrated numerically by using the trapezoidal rule, also with $10^4$ discrete intervals. This is carried out for an initial, tentative value $\sigma^*$ (normally inferred from the experimental curve), which is then adjusted to provide the best representation of data.

One should underline that the present calculation takes into account the variations of $\dot{\gamma}^<(\sigma)$ and $\dot{\gamma}^>(\sigma)$ (i.e. spatial variations) in each band. Hence, it differs notably from current assumptions made in several works [6,34,58,72,73]. In these studies, $\dot{\gamma}^<$ and $\dot{\gamma}^>$ are taken as the limits $\dot{\gamma}_1$ and $\dot{\gamma}_2$ of the plateau $\sigma^*$, hence are the constants entering the "lever rule" $e\bar{\dot{\gamma}} = e_1\dot{\gamma}_1 + e_2\dot{\gamma}_2$, where $e_1$ and $e_2$ are the respective band thickness, $e = R - \kappa R$ being the gap width, and $\bar{\dot{\gamma}}$ being the "measured" shear rate. Such constant values are not observed in measured velocity profiles (see, for example, Fig.2 in [6]). Moreover, it is evident that the larger the gap thickness $e$, the higher the differences from constant values of $\dot{\gamma}$.

## 5 - Examples of application

This section illustrates the applicability of the SN-model to represent rheometric data of shear-banding fluids. The first step consists in confronting Eq.(2) to data $\sigma$ vs. $\dot{\gamma}$ from real systems. For this purpose, experimental data of WMS published in the literature are considered: CTAB [6] and CPCl-NaSal [7]. These data, which were reported as $\sigma_{\kappa R}$ vs. $\dot{\gamma}_{ng}$, are presented in Fig.7 as $\sigma_{\kappa R}$ vs. $\Omega = \dot{\gamma}_{ng}(1-\kappa)/\kappa$, by using the corresponding values of $\kappa$ (0.94 and 0.96, respectively).

Applying Eq.(2) firstly requires the estimation of accurate values $\sigma_{\kappa R}$ vs. $\dot{\gamma}_{\kappa R}$ from data $\sigma_{\kappa R}$ vs. $\Omega$. Thus we selected intervals of experimental data of Fig.7 where the fluid is monophasic only. The low and high shear zones were analysed independently one another to find $\dot{\gamma}_{\kappa R}$ by means of Eq.(6). Results are presented in Fig.8 (symbols).

Afterwards, Eq.(2) is adjusted to data $\sigma_{\kappa R}$ vs. $\dot{\gamma}_{\kappa R}$ (full lines in Fig.8), leading to the following parameter values:

CPCl/NaSal : $\eta_0 = 63.6$ Pa.s , $\eta_\infty = 0.64$ Pa.s , $t_c = 32$ ms ;

CTAB : $\eta_0 = 7.8$ Pa.s , $\eta_\infty = 25$ mPa.s , $t_c = 2.7$ ms.

It is observed that the model describes satisfactorily the flow curve in the full range of shear rates, by predicting an intermediate multivalued zone.

In order to cross-check these results, and having the parameters $\eta_0$ , $\eta_\infty$ , and $t_c$ that characterize the fluid, we finally carried out the direct calculation, defined in Sec. 4.3. That is, Eqs.(9a-c)-(2) were calculated numerically, and the resulting function $\Omega(\sigma_{\kappa R})$ was matched to the respective experimental curve. This is actually done in Fig.9, where full lines are the numerical predictions with the adjusted values of $\sigma^*$ indicated in the figure caption. A remarkable agreement is observed in the full range of experimental data. Furthermore, it is worthy of note that the values of $\sigma^*$ obtained are in close agreement with those previously reported [6,7].

# 6 - Concluding remarks and perspectives

A phenomenological (structural) model has been introduced in order to improve rheological modelling of complex fluids under shear banding conditions. In contrast with the majority of other minimal models already developed in the literature for shear banding description, this model has a "structural origin" (a kinetic description which accounts for shear induced changes of the internal structure of the material) which seems to bring forwards a better physical basis. Simplest assumptions on structure kinetics lead to a Strucural Minimal (SM-) model. Existence of a flow curve, multivalued in shear rate −a necessary requirement for the presence of shear banding−, has been found depending only on the ratio of the limiting viscosities at very high and very low shear.

Under controlled (increasing) stress conditions, the SM-model predicts a *top-jumping*-like behaviour with a smooth transition to an apparent plateau. The later cannot represent the true plateau associated to shear banding since such predictions concern nonequilibrium responses as those that could result from too fast measurements. Similarly, for decreasing stress, a *bottom-jumping* is predicted, forming with top-jumping a loop which resembles a hysteresis cycle. However, such findings have not been further discussed in the paper since many works have demonstrated a lack of stability in the vicinity of the stress plateau if non-local effects are absent from the model used, as it is the case in this work.

Prediction of given rheometric data of shear-banding fluids has been a test for this constitutive model of the flow curve. It may be remarked that the success of calculations suggested relies on both the introduction of a suitable model for the fluid, and the adequate computation of the shear rate function in the rheometric cell. Indeed, determining the true shear rate attained in Couette flows is a non-trivial task, except if the gap between cylinders is very small. When fluids presenting strong shear-thinning behaviour are studied, an inappropriate estimation of the shear rate leads to considerable errors, notably when two phases coexist in the flow domain. In this work, although the absence of non-local terms in the SM-model impedes the stress plateau value $\sigma^*$ to be determined, successfully testing of the model has been nevertheless obtained by adding $\sigma^*$ to the three model variables $\eta_\infty$, $\eta_0$ and $\dot{\gamma}_C$ to be adjusted by data fitting.

A crucial aspect in complex fluids like WMS, and more generally systems exhibiting shear banding phenomena, is the modelling of unsteady rheometric data, specially stress relaxation after the sudden inception of a given shear rate. A stress overshoot is observed, followed by dumped oscillations and either or not Maxwellian relaxation to the steady state [4,5,33]. Theoretical descriptions of these experiments are rather demanding, as different time-dependent phenomena are involved, mainly fluid microstructure relaxation and fluid viscoelasticity, even the transient characteristics of the apparatus in some studies [74]. The structural model from which Eq.(2) derives also includes shear induced kinetic processes involved in structural changes (see [53]). A nonlinear Maxwell model based on Eq.(2) allowed to interpret unsteady curves on the basis of such kinetic processes [54,75], hence to predict transient responses like those observed in WMS. This work is in progress.

Future work will be devoted to investigate if the SM-model predictions after including non-local terms will make possible selecting the $\sigma^*$-value and further would provide some new physical argument in the debate on the origin of shear banding (i.e. formation and robustness of the stress plateau) in favour of either a mechanical instability or a shear-induced phase transition [34]. Moreover, these model predictions will be aimed to be compared with those from Johnson-Segalman model, which is nowadays the most employed constitutive relation to describe shear-banding fluids [39,43–47,68,72], as well as the so-called BMP model [32].

**Acknowledgements:** C. B. acknowledges CONICET and SEPCYT-FONCyT, Argentina, for the financial aid received.

**Figure captions**

**Figure 1:**

Curve of rheometric data (arbitrary drawing) typically found in WMS studied by Couette flow. The insets are highly schematic representations of the solution structure in different shear rates zones. At a given shear stress, a new fluid phase (*band*) develops as micelles align in the flow direction.

**Figure 2:**

Shear stress as a function of shear rate (arbitrary units) for shear-banding fluids.

**Figure 3:**

A – Reduced stress $\Sigma$ *vs.* reduced shear-rate $\Gamma$ from Eq.(2) for different values of the limiting viscosity ratio $\alpha = \eta_0/\eta_\infty$. The non-monotonicity disappears if $\alpha < 81$.

B – Reduced viscosity $H = \eta/\eta_\infty$ *vs.* reduced shear-rate $\Gamma$ for the same values of $\alpha$.

(from top to bottom : $\alpha =$ 1000; 200; **81**; 50; 10; 1; 0.1; 0.01).

**Figure 4:**

A – Reduced stress $\Sigma$ *vs.* reduced shear-rate $\Gamma$, defined in Eq.(3), for different values of the limiting viscosities ratio $\beta = \eta_\infty/\eta_0$. The non-monotonicity disappears if $\beta < 81$.

B – Reduced viscosity $H = \eta/\eta_\infty$ *vs.* reduced shear-stress $\Sigma$ for the same values of $\beta$.

C – Reduced viscosity $H = \eta/\eta_\infty$ *vs.* reduced shear-rate $\Gamma$ for the same values of $\beta$.

(from top to bottom : $\beta =$ 1000; 200; **81**; 50; 10; 1; 0.1; 0.01).

**Figure 5**:

A- Reduced stress *vs.* reduced shear rate under controlled stress conditions, showing a top-jumping –like behaviour (△) (Stress increment : $d\Sigma = 4.10^{-2}$), in comparaison with the flow curve (····) under controlled shear rate conditions.

B- Details of Fig.5A: the linear scale for $\Gamma$ allows one to appreciate the increasingly successive jumps $d\Gamma$.

Model parameters: $\eta_0 = 10^3$ Pa.s, $\eta_\infty = 0.1$ Pa.s, $t_c = 1$ ms ($\dot{\gamma}_C = 10^3$ s$^{-1}$) $\Leftrightarrow$ $\chi = 0.01$; $\sigma_c = 10^2$ Pa

**Figure 6:**

Idem Fig.5, showing top- and bottom- jumpings, leading to a hysteresis cycle.

**Figure 7:**

Angular velocity as a function of controlled shear stress for different WMS. Symbols represent two examples of experimental data reported in the literature: CTAB [6], and CPCl/NaSal [7].

**Figure 8:**

Shear stress as a function of shear rate, corresponding to the systems presented in Fig.7. Symbols are the values obtained from experimental data $\sigma_{\kappa R}(\Omega)$ in the shear zones where the fluid is monophasic (see text for details). Full lines are the predictions of Eq.(2), with the parameter values reported in the text.

**Figure 9:**

Shear stress as a function of angular velocity for different WMS. Symbols represent two examples of experimental data reported in the literature: CTAB [6], and CPCl/NaSal [7]. Full lines are the predictions of Eqs.(9a-c)-(2), with the values of $\eta_0$, $\eta_\infty$, and $t_c$ reported in the text. In addition, $\sigma^* = 62$ Pa for CPCl/NaSal, and $\sigma^* = 32.5$ Pa for CTAB.

Figure 1  Quemada and Berli 2008

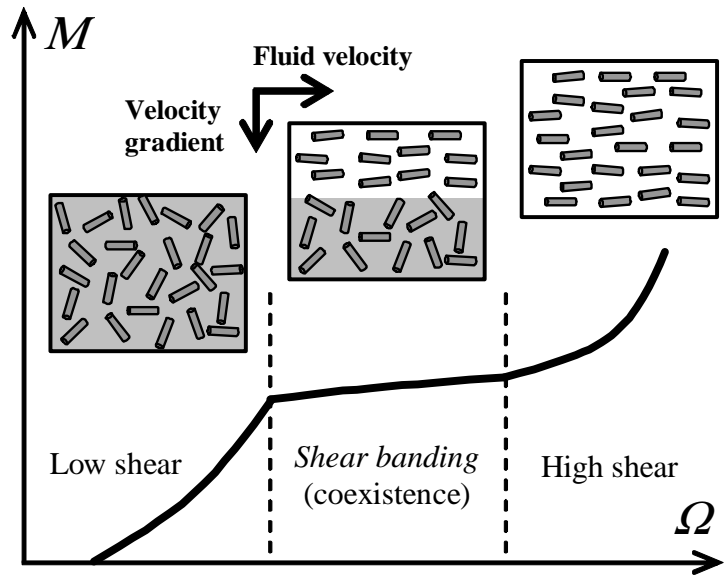

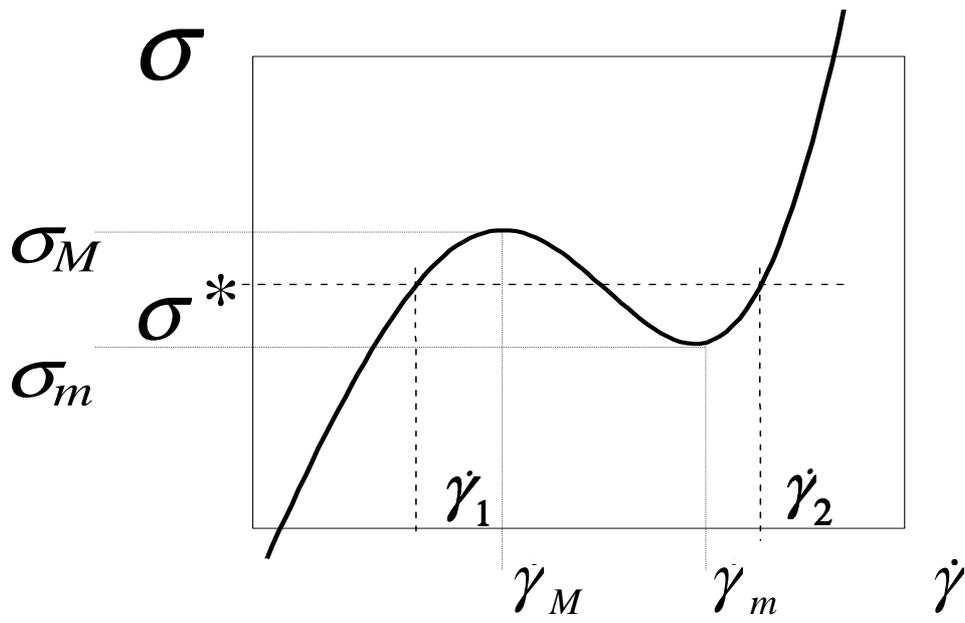

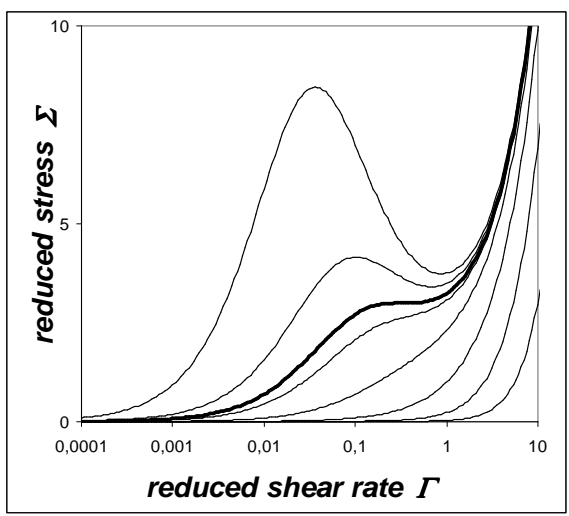
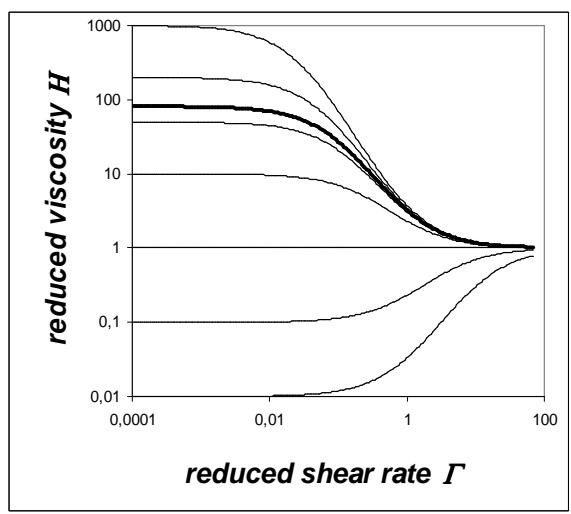

A  B

Figure 4 Quemada and Berli 2008

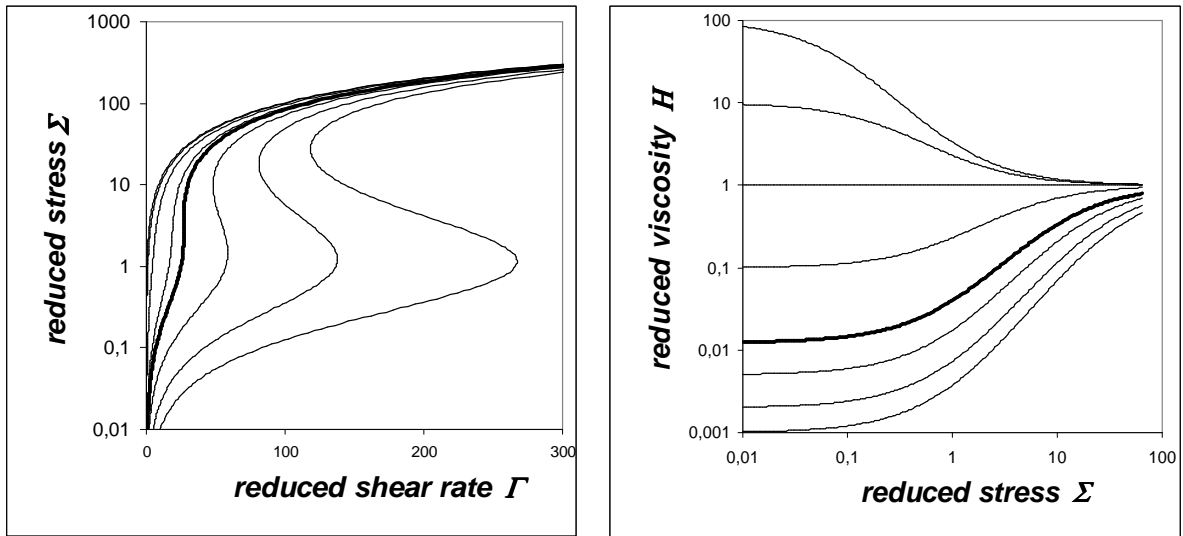

A

B

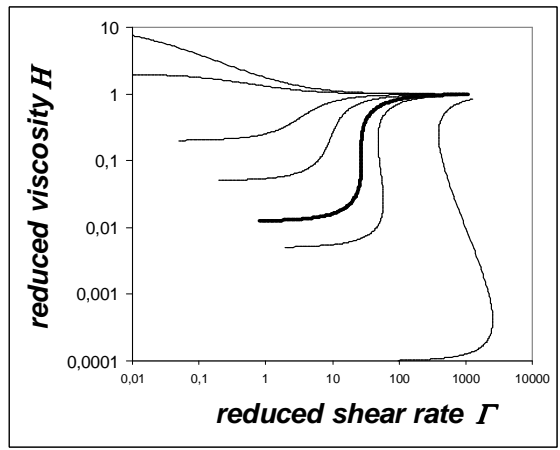

C

Figure 5                                    Quemada and Berli 2008

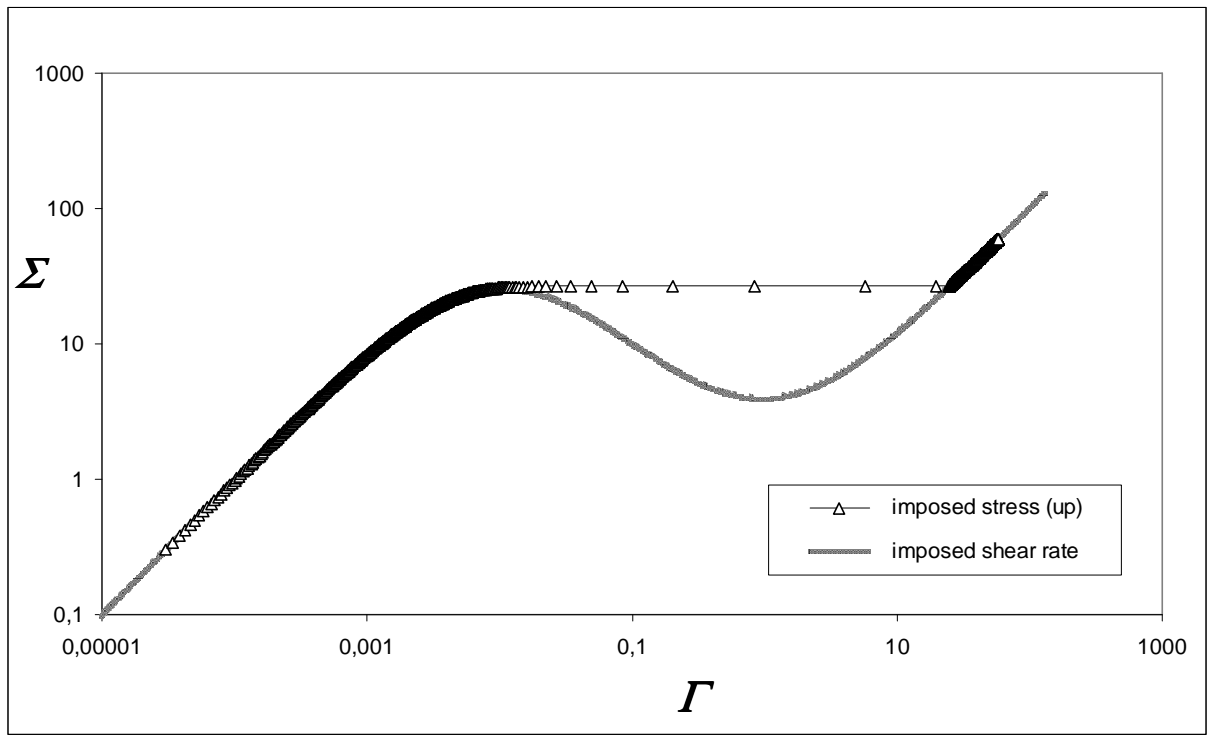

A

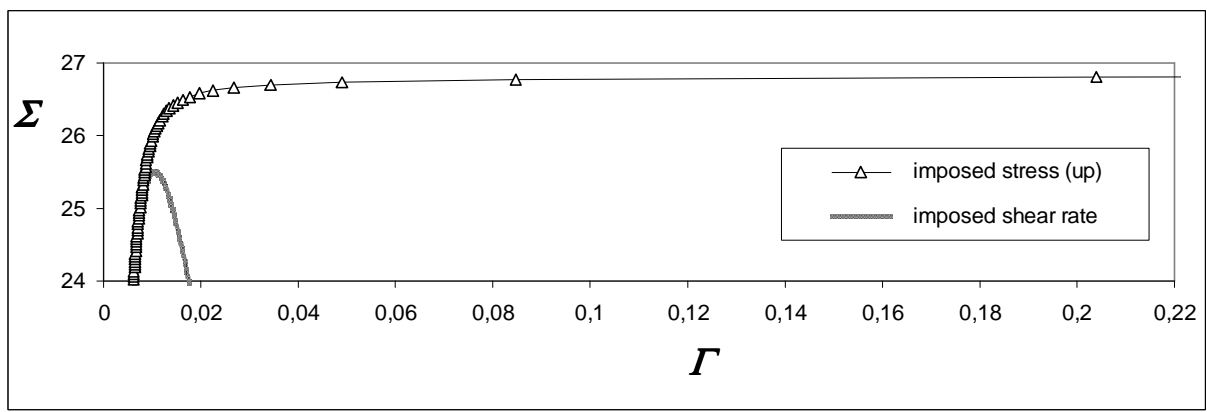

B

Figure 6                                                     Quemada and Berli 2008

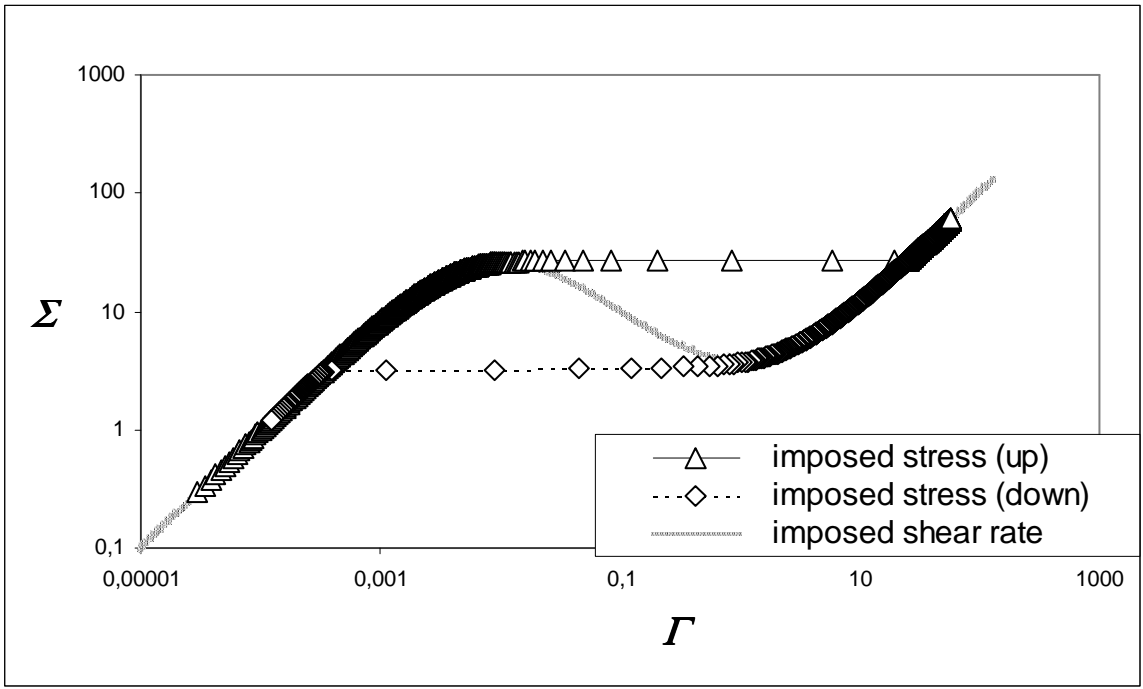

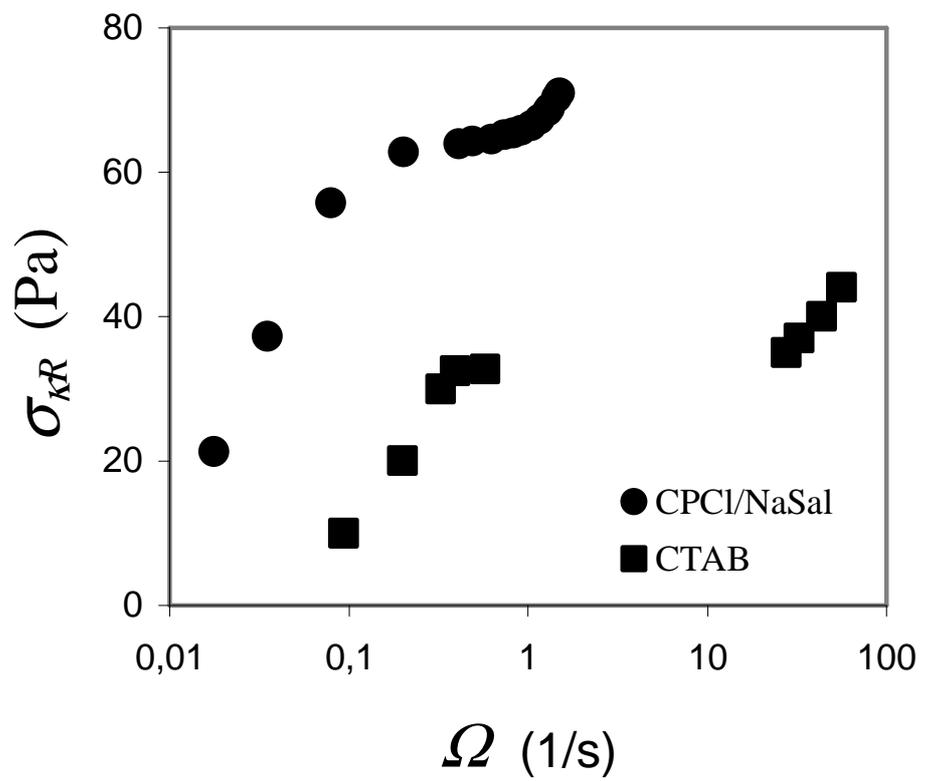

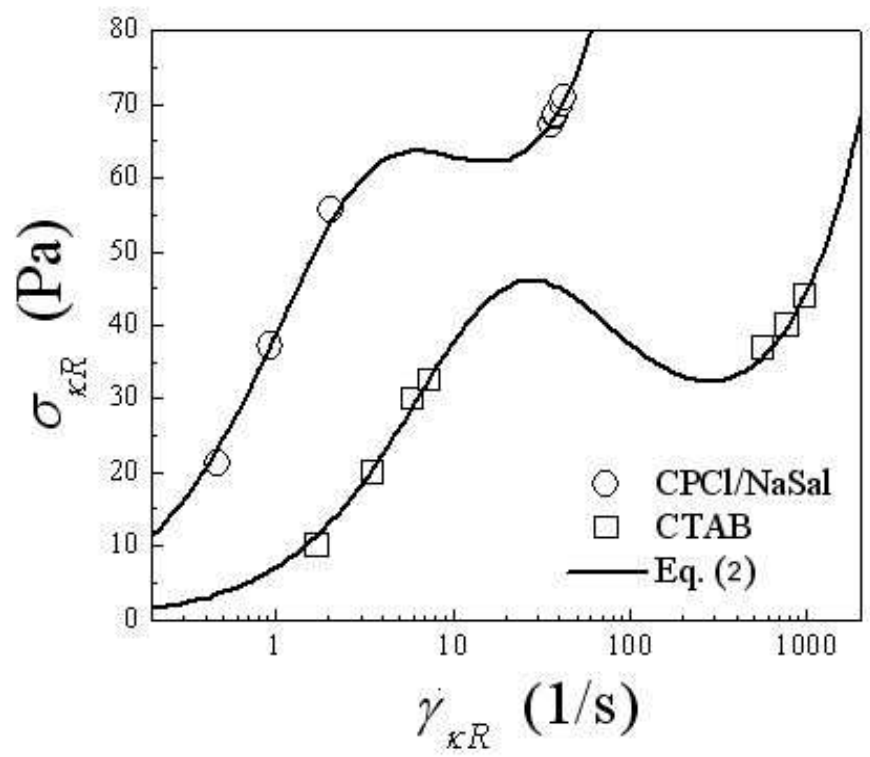

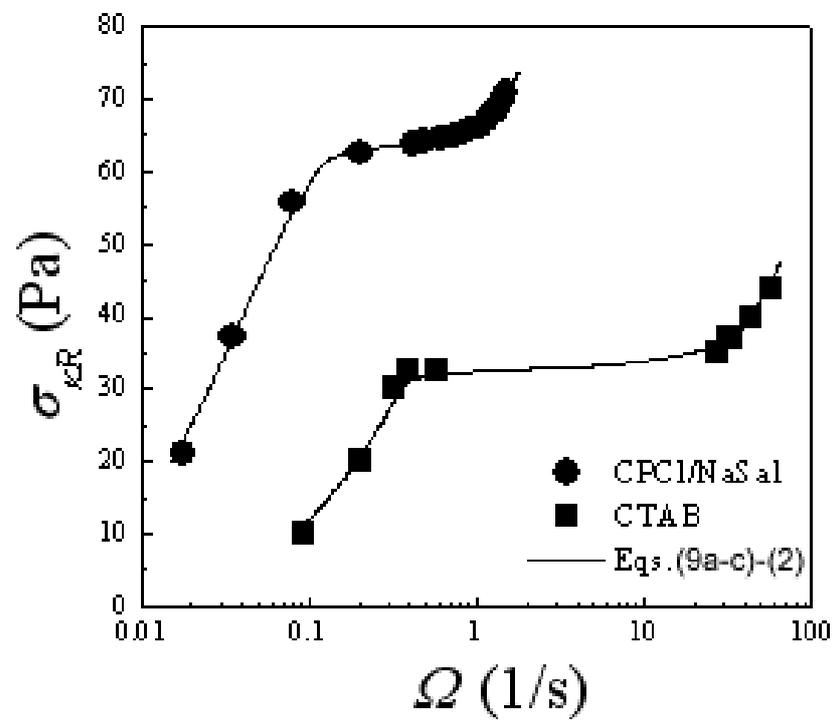